\documentclass[A4,aps,twocolumn,notitlepage,superscriptaddress,nofootinbib,floatfix]{revtex4-2}
\usepackage{amsmath,amssymb,amsfonts}
\usepackage{bbm,bm}
\usepackage[utf8]{inputenc}
\usepackage{amsmath,amsfonts,amssymb,graphicx,subfigure}
\usepackage{slashed,color}
\usepackage{multirow}
\usepackage{lineno}
\usepackage{ulem}
\usepackage{verbatim}
\usepackage[svgnames]{xcolor}
\usepackage{hyperref}

\definecolor{codegreen}{rgb}{0,0.6,0}
\definecolor{codegray}{rgb}{0.5,0.5,0.5}
\definecolor{codepurple}{rgb}{0.58,0,0.82}
\definecolor{backcolour}{rgb}{0.95,0.95,0.92}

\usepackage{listings}
\lstdefinestyle{mystyle}{
    backgroundcolor=\color{backcolour},
    commentstyle=\color{codegreen},
    keywordstyle=\color{magenta},
    numberstyle=\tiny\color{codegray},
    stringstyle=\color{codepurple},
    basicstyle=\ttfamily\footnotesize,
    breakatwhitespace=false,
    breaklines=true,
    captionpos=b,
    keepspaces=true,
    numbersep=5pt,
    showspaces=false,
    showstringspaces=false,
    showtabs=false,
    tabsize=2
}
\usepackage{hyperref}
\hypersetup{colorlinks=true, linkcolor=blue, citecolor=ForestGreen,
filecolor=magenta, urlcolor=ForestGreen,
  pdftitle={NAMES [arXiv:2507.xxx]},
  pdfauthor={NAMES},
  pdfsubject={Subject},pdfcreator={Creator},pdfproducer={Producer},
  pdfkeywords={Data Science Methods}{Demonstration}{arXiv}{Social Media}
}
\graphicspath{{Figs/}}

\definecolor{magenta}{HTML}{FF00FF}
\definecolor{cornflowerblue}{HTML}{6495ED}
\definecolor{turquoise}{HTML}{40E0D0}

\definecolor{darkgreen}{rgb}{0.0, 0.2, 0.13}
\definecolor{darkmagenta}{rgb}{0.55, 0.0, 0.55}
\definecolor{amber}{rgb}{1.0, 0.6, 0.0}

\newcommand{\confirm}[1]{{\color{black}#1}}
\newcommand{\orcid}[1]{\,\href{https://orcid.org/#1}{\includegraphics[width=9pt]{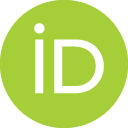}}}
\newcommand{\orcidRR}{0000-0002-3316-2175} 
\newcommand{\orcidSD}{0000-0002-2447-5989} 

\begin{document}
\leftline{}
\rightline{IFJPAN-IV-2025-14, COMETA-2025-30}

\title{Are arXiv submissions on Wednesday better cited?\\
Introducing Big Data methods in undergraduate courses on  scientific computing}

\author{St\'ephane Delorme\ \orcid{\orcidSD}}
\email{stephane.delorme@us.edu.pl}
\affiliation{Institute of Physics -- University of Silesia, Katowice, Poland}

\author{Leon Mach}
\affiliation{Faculty of Physics -- 
University of Warsaw, Pasteura 5, Warsaw, 02-093, Poland}

\author{Hubert Paszkiewicz} 
\email{hubolaf@gmail.com}
\affiliation{Faculty of Physics -- 
University of Warsaw, Pasteura 5, Warsaw, 02-093, Poland}

\author{Richard Ruiz\ \orcid{\orcidRR}}
\email{rruiz@ifj.edu.pl}
\affiliation{Institute of Nuclear Physics -- 
Polish Academy of Sciences {\rm (IFJ PAN)}, ul. Radzikowskiego, Krak{\'o}w, 31-342, Poland}


\begin{abstract}
Extracting
information from
big data sets, both real and simulated,
is a
modern hallmark of 
the physical sciences.
In practice, 
students face barriers
to learning 
 ``Big Data'' methods
in undergraduate physics \& astronomy curricula.
As an attempt to alleviate some of these challenges,
we present 
a simple, farm-to-table data analysis pipeline
that can collect, process, and plot data from the
\confirm{800k} entries 
common to
the \textsc{arXiv} preprint repository
and the bibliographical  database \textsc{inSpireHEP}.
The pipeline employs contemporary research practices 
and can be implemented using
open-sourced \textit{Python} libraries
common to undergraduate courses on
\textit{Scientific Computing}.
To support the use such pipelines in classroom contexts,
we make public an example implementation,
authored by two  undergraduate physics students,
that runs on off-the-shelf laptops.
For advanced students,
we discuss applications of the pipeline,
including for online DAQ monitoring 
and commercialization.

\end{abstract}
\date{\today}
\maketitle


\section{Introduction}
\label{sec:intro}

Plotting data and establishing (or falsifying) correlations with statistical confidence
are among the most basic
skills of a physical scientist.
So core to physics \& astronomy,
the topics are introduced
as early as high school,
often in labs aimed at extracting the gravitational constant on Earth's surface.
In practice,
many students carry out these first analyses
with spreadsheet software 
(e.g., \textit{Excel}),
non-graphing calculators,
or just pen and paper.

While the core principles are the same,
contemporary analyses in physics \& astronomy
research are obviously much more sophisticated.
For many disciplines,
such as astrophysics, nuclear physics
and particle physics,
analyses of real data and 
simulated data
fall into the realm of  ``Big Data.''
Datasets are too big and/or too complex
to be assessed with multipurpose tools like \textit{Excel}.

For example:
Despite historically 
being a standard tool in industries
that employ former STEM 
students (e.g., finance),
\textit{Excel} is
limited
to
$2^{20}\sim\mathcal{O}(10^6)$ data entries (rows)
and $2^{14}\sim\mathcal{O}(10^4)$ data attributes (columns).
In comparison,
experiments at CERN's Large Hadron Collider
detect more than 
$10^{16}$ proton collisions,
or about $10^{10}$ \textit{Excel}
spreadsheets,
per year~\cite{ATLASCollaboration:2012ilu,Contardo:2015bmq}.
Similarly, Data Release 1 of the Dark Energy Spectroscopic Instrument
contains 18.7 million objects, or about 18
spreadsheets~\cite{DESI:2025fxa}.
Even a voter registry for a large city
can exceed $10^{6}$ entries.
In other words, 
the skills and tools needed for 
data-heavy sectors 
inside and outside academia 
now exceed those taught 
in introductory laboratories.

Efficiently analyzing big data sets 
in a timely manner requires training,
typically at the PhD level,
in fast compiled languages, like \textit{C++},
and high-level scripting languages, like \textit{Python}.
It requires constructing and interfacing
analysis pipelines.
It requires managing memory in
loops over entries.
It requires knowledge of software libraries
(e.g., \texttt{matplotlib}, \texttt{scipy}, \texttt{pandas})
in addition to physics and statistics.
So on and so forth.
However,
to carry out undergraduate research projects
and to be competitive for graduate programs,
prospective physicists and astronomers
are increasingly expected to already have
such  skills \textit{at the bachelor's 
level}~\cite{Mulvey_2025_Employment,Mulvey_2025_Skills,Mulvey_2025_AI}.

Certainly, 
where available, 
Big Data practices
are being introduced
in
undergraduate courses on
scientific computing
and undergraduate research opportunities.
In practice though,
students face barriers
to learning these skills~\cite{ShahKnaub:2024}.
Practical challenges include:
a lack of 
example scripts that are 
scalable to large datasets
and
comprehensible to students 
outside of
computer science;
limited access to computing resources
(e.g., many-processor computing clusters);
and limited access to large datasets themselves.
Simply put,
the jump from using
\textit{Excel} in first-year laboratories
to using research-level 
analysis frameworks,
like \texttt{Astropy}~\cite{Astropy:2013muo}
or
\texttt{ROOT/pyROOT}~\cite{Brun:1997pa,Antcheva:2009zz},
in third- and fourth-year studies
is difficult.

To help bridge the growing gap~\cite{ShahKnaub:2024}
between the computational skills
that undergraduate students are explicitly taught
and
those they are expected to
just know\footnote{In this sense,
computational skills are part of the
``hidden curriculum'' in physics \& astronomy.
See Ref.~\cite{aaptGuestEd} and references therein,
including the AAPT \textit{Recommendations for
Computational Physics
in the Undergraduate Physics Curriculum}.},
we explored 
whether a light-weight,
analysis framework could be developed
that is also suitable for
undergraduate  
 physics \& astronomy courses.
 Furthermore, we explored 
 whether this was possible without requiring 
specialized or advanced physics knowledge.
The outcome was positive.

\begin{figure}[t!]
  \includegraphics[width=\columnwidth]{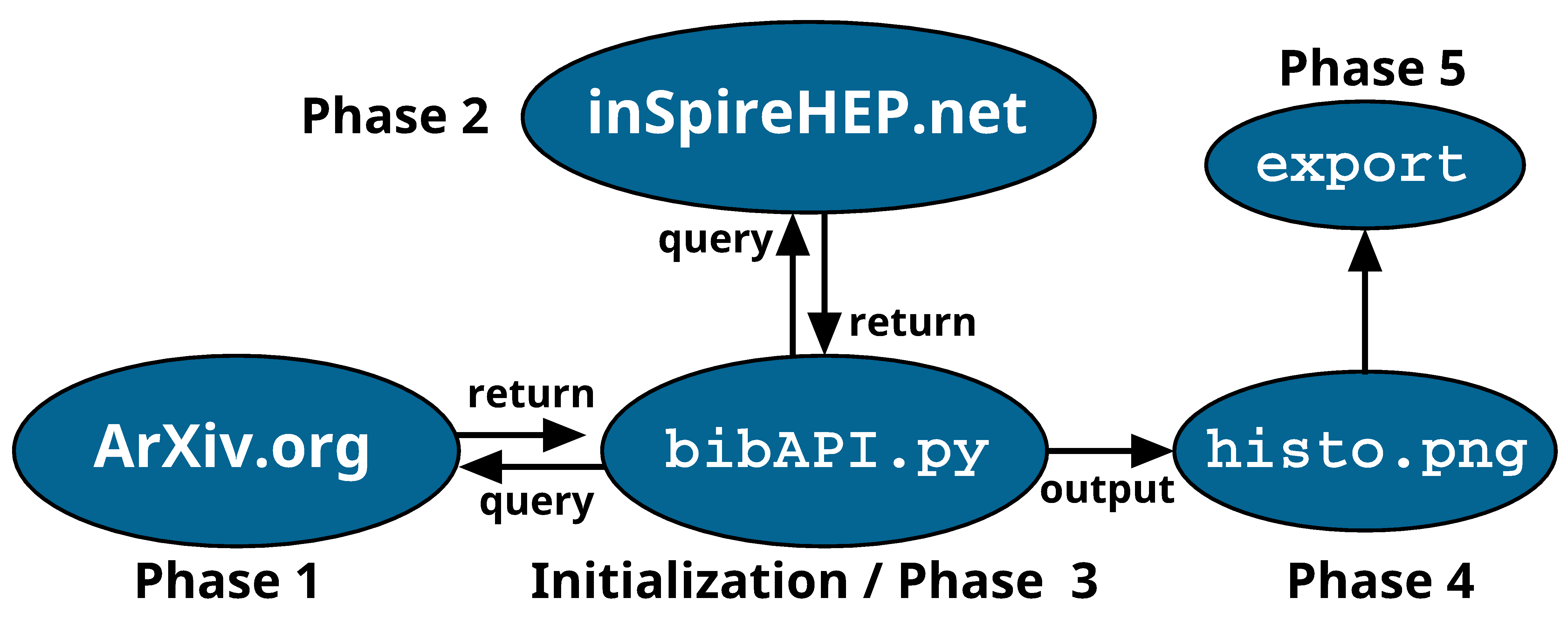}
  \caption{Illustration of a ``farm-to-table''
  data analysis pipeline, including 
  pipeline initialization   (Phase 0),
  data queries to the \textsc{arXiv} (Phase 1),
  data queries to \textsc{inSpireHEP} (Phase 2),
  processing/filtering data (Phase 3),
  plotting/analyzing data (Phase 4),
  exporting output (Phase 5).
  }
\label{fig:apiFlow_arXivInSpire}
\end{figure}

As part of a four-week
summer 
research 
program\footnote{\label{foot:ppss}Program details available at   
\href{https://ppss.ifj.edu.pl/}{https://ppss.ifj.edu.pl}.},
two mid-level undergraduate physics students
successfully built
a ``farm-to-table'' analysis pipeline
employing methods from Big Data.
As illustrated in Fig.~\ref{fig:apiFlow_arXivInSpire},
the pipeline
collects/harvests, processes, {and} plots
bibliometric data for entries 
common to
the \textsc{arXiv} preprint repository 
and
the \textsc{inSpireHEP} bibliographical  database,
which are both publicly accessible.
(The pipeline itself mirrors those used 
in high energy physics and other fields.)

The key to the pipeline's design,
and therefore usefulness 
in overcoming practical challenges,
is the use of Application Program 
Interfaces\footnote{For non-experts, APIs 
are a set of call-and-response
rules, i.e., a protocol,
between computer platforms.
Calling an API entails calling 
a base URL extended by markup text.
The receiver interprets the full URL and
returns the call, 
often as \texttt{txt} or \texttt{xml}.} (APIs).
APIs are prevalent in data- and web-based industries,
especially social media (e.g., ``click to share'' buttons).
They also have critical uses in 
experiments,
such as real-time online data acquisition monitoring
and telescope alert systems.

Starting from the APIs for the \textsc{arXiv} and \textsc{inSpireHEP},
the students implemented the pipeline into a set of \textit{Python} scripts
using popular, open-sourced libraries
(\texttt{numpy}, \texttt{matplotlib}, etc.).
The precise choice of these databases 
is somewhat arbitrary and due to familiarity,
but is nevertheless illustrative, 
as discussed in Sec.~\ref{sec:method}.
Remarkably, 
each phase of the pipeline 
required only $\mathcal{O}(100)$ lines of code
and ran on personal, off-the-shelf laptops.

Due to the granularity of the data
and the largeness of the databases, 
students were able to explore 
rather {creative} correlations 
with statistical confidence.
For example:
\confirm{using around 330k manuscripts
from particle and nuclear physics over a 34-year period,}
students found no correlation
between citation count and
\textsc{arXiv} submission date,
contrary to popular lore.
To support the use of the pipeline in coursework,
an example implementation by the students 
is freely available at the repository
\begin{center}
    \href{https://gitlab.cern.ch/riruiz/public-projects/-/tree/master/BibAPI}{https://gitlab.cern.ch/riruiz/public-projects/-/tree/master/BibAPI} 
\end{center}
Throughout the text, we refer to this template as \texttt{BibAPI}.

In Sec.~\ref{sec:method} we give an overview of 
our ``farm-to-table'' analysis pipeline.
In Sec.~\ref{sec:implementation}, 
we discuss its implementation 
by the undergraduates into \texttt{BibAPI}.
Example findings are presented in Sec.~\ref{sec:examples},
while pipeline extensions and applications
are discussed in Sec.~\ref{sec:extensions}.
We conclude in Sec.~\ref{sec:conclusion}.
Our computational setup and some supplemental results
are given in the appendices.

\section{Overview of analysis pipelines}
\label{sec:method}

In physics \& astronomy research,
the general objective of 
data-analysis pipelines is to
carry out the following operations:
(i) access data from one or more databases,
(ii) organize and tabulate a subset of the bulk data,
(iii) analyze the subset(s),
and
(iv) plot and record the results.
Data may be real or simulated,
signal-like or background-like.
Subsets are built
by filtering according
to data attributes.
Analyzing and plotting may occur iteratively.

When introducing such pipelines into
 undergraduate physics \& astronomy courses,
we propose relaxing the expectation
that students analyze data from
 real experiments or simulations.
Even if large data sets are available,
students may not have sufficient
training to comprehend 
their complexity,
particularly first- and second-year students.
An adverse consequence is that some students
obtain erroneous or no results
because they have not yet 
taken upper-level courses,
e.g., special or general relativity.
Learning challenges are compounded if pipelines are
{simultaneously} being used to introduce 
error analysis.

In lieu of a physics data set,
we advocate using public
 socioeconomic and bibliometric data.
While broad interpretation of such data requires care,
individual data attributes (columns)
can be straightforward to comprehend.
In addition to demonstrating
how physics skills are applicable outside the field,
it retains an interest factor as such
public data\footnote{The City of Chicago's Data Portal, for example,
offers a dataset of reported crimes since 2001.
Presently, the dataset contains \confirm{30} data attributes (columns)
for more than \confirm{8.3M} entries (rows) and is available at 
\href{https://data.cityofchicago.org/Public-Safety/Crimes-2001-to-Present/ijzp-q8t2/about_data}
{https://data.cityofchicago.org/Public-Safety/Crimes-2001-to-Present/ijzp-q8t2/about\_data}.} 
steers public policy and discourse.

As a proof-of-principle,
we focus on analyzing bibliometric data
of manuscripts in particle and nuclear physics
submitted to the \textsc{arXiv} 
 \confirm{from 1991 to March 2025}.
This constitutes about \confirm{330k} entries
out of the \textsc{arXiv}'s \confirm{2.8M+}
entries across all disciplines.
For each  manuscript,
its bibliometric record
was extracted from \textsc{inSpireHEP}.

The workflow of our
``farm-to-table'' pipeline is
illustrated in Fig.~\ref{fig:apiFlow_arXivInSpire}.
We start by querying the 
\textsc{arXiv} through its API (\textbf{{Phase 1}}).
The query returns a list of manuscripts matching our 
search criteria
(i.e., specific subject categories 
over a particular date range).
Each manuscript entry contains identifying data and submission metadata.
For each entry in the list, 
a query is made to \textsc{inSpireHEP} via its own 
API~\cite{Moskovic:2021zjs}
(\textbf{{Phase 2}}).
Each  secondary query returns blibliometric data,
such as authorship,
citation count,
etc.
Data are then 
filtered 
and composite quantities (averages and ratios) 
are calculated (\textbf{{Phase 3}}).
Finally,
results are plotted (\textbf{{Phase 4}})
and
possibly exported to online platforms (\textbf{{Phase 5}}).

\section{Pipeline Implementation}
\label{sec:implementation}


We now describe in fuller detail
the pipeline of 
Fig.~\ref{fig:apiFlow_arXivInSpire} 
as implemented into \texttt{BibAPI}.
%
%
For exporting results to online platforms 
via APIs (\textbf{Phase 5}), 
see Sec.~\ref{sec:extensions_socials}.

We note that the students' \texttt{bibAPI.py} template
has been modularized by the senior authors
to allow each \textbf{Phase} to run independently.
By embedding each \textbf{Phase} into a function
(an example of ``functional programming'')
students and instructors can 
study, modify, and extend individual sequences
without repeating calls to online databases.
Information is shared between \textbf{Phases}
by passing an instance of 
the \texttt{APIQueryInputs()} class,
which retains ``state'' information 
(an example of ``object oriented programming'').
While modularization can improve readability 
(about \confirm{100-250} lines per phase)
and facilitate lesson planning,
it also leads to ``code bloat.'' 
In total, the script is about 
\confirm{1000 lines}, 
including comments,
checks and error handling,
and \textbf{Phase 5}.

\subsection{Students' Experiences and Backgrounds}
\label{sec:implementation_prior}

The students started work on the pipeline in 
Summer 2024 as part of a 4-week Research Experience for 
Undergraduate program (see footnote \ref{foot:ppss}).
Guided by a postdoctoral researcher,
they developed the pipeline at a pace of roughly 
one phase per week. 
As a  cross check, 
each student implemented their own version of the pipeline.

At the conclusion of the program, 
students successfully present 
preliminary versions of 
results shown in Sec.~\ref{sec:examples}.
Due breaks and term-time obligations,
work was paused until Spring 2025.
The students' works were then merged
and patched for bugs by \textbf{Student 2}
and \textbf{Faculty 1}.

Prior to Summer 2024 
\textbf{Student 1} had just completed their 
\confirm{second}  year in a physics bachelor program. 
\textbf{Student 1} briefly majored in computer science
in their first year of studies.
At the start of the project 
\textbf{Student 2} had similarly just completed their 
\confirm{second} year in another physics bachelor program. 
\textbf{Student 2} was introduced to 
\confirm{plotting} with \textit{Python}
in a course on basics of programming languages used in modern physics 
but had not been introduced to more advanced applications of \textit{Python}.

\textbf{Postdoc 1} already had prior experience with \textit{Python} and other programing languages, obtained during their PhD work 
and first postdoctoral appointment. The usage of APIs was however completely new and had to be learned. 
\textbf{Faculty 1} is a 
theoretical and computational 
high energy physicist and develops
simulation software.

\subsection{Implementation  into \texttt{BibAPI}}
\label{sec:implementation_script}

\textbf{Phase 0 (Initialization)}
occurs when the script is 
run from the terminal or command prompt. 
During this phase
\textsc{arXiv} query parameters are 
set either to their default values 
(\texttt{apiInputs = APIQueryInputs()}),
or the those specified in the terminal 
(e.g., \texttt{apiInputs.cats = sys.argv[3:]}). 
Allowed inputs are the  
start/end dates of the 
\textsc{arXiv} 
query (\texttt{YYYY-MM-DD} format)
and 
subject categories 
(\texttt{hep-th}, \texttt{nucl-ex}, etc.).
Paths for 
files written to disk are also set here.
(These must be modified manually 
in \texttt{\_\_main\_\_}).
In the \texttt{\_\_main\_\_} sequence, 
\textbf{Phase} tasks are carried out 
by the functions \texttt{doPhaseN(\dots)}, 
whose inputs are an instance of \texttt{APIQueryInputs}.

\textbf{Phase 1 (ArXiv Query)}
fetches data from the URL 
\href{http://export.arxiv.org/api}{export.arxiv.org/api}. 
This 
is the \textsc{arXiv}'s  subdomain 
that hosts its API and 
functions as a portal to its database.
A list of manuscripts and submission records
can be obtained by appending the URL
by query parameters in a prescribed format
(\texttt{/query?search\_query=\dots}).
A query returns an \texttt{xml} packet in the \texttt{atom} format. 

Inputs passed to \texttt{doPhase1(\dots)} are used to build a set of ``micro queries,''
where the specific date range of the full query is divided into three-month windows
(\texttt{generate\_date\_ranges(\dots)}).
Each ``micro URL'' is then opened (\texttt{urllib.request.urlopen(\dots)}).
Returned \texttt{xml} packets
are parsed and 
saved to a text file
(\texttt{APIQueryInputs.idPath[0]})
with tab-separated values.
To prevent denial-of-service attacks,
a short break is taken (\texttt{time.sleep(2)}) after multiple queries.

\textbf{Phase 2 (InSpireHEP Query)} fetches data from the URL
\href{https://inspirehep.net/api}{inspirehep.net/api}.
This is \textsc{InSpireHEP}'s subdomain hosting its own API.
Appending this base URL by \texttt{/arxiv/}
and 
a manuscript's \textsc{arXiv} 
identifier\footnote{For example: \href{https://inspirehep.net/api/arxiv/0812.3578}{https://inspirehep.net/api/arxiv/0812.3578}.}
will return the manuscript's 
bibliometric record in \texttt{json} format.

The function \texttt{doPhase2(\dots)} starts by loading the 
list of manuscript identifiers (\texttt{APIQueryInputs.idPath[0]})
generated by \texttt{doPhase1(\dots)}.
For each manuscript in the list, 
a query URL is constructed, called, and read 
(\texttt{json.loads(urllib.request.urlopen(\dots).read())}).
Records are parsed and saved to a text file with tab-separated values
(\texttt{APIQueryInputs.dataPath[0]}).
As in \textbf{Phase 1}, 
the code pauses 
after multiple queries.

Sometimes information is not known by \textsc{InSpireHEP} 
(e.g., publication details of an unpublished paper). 
In these instances, ``\texttt{Not-Given}'' is recorded
as 
blank spaces can result in 
data misattribution.
Some \textsc{arXiv} papers, 
e.g., those withdrawn,
do not have an \textsc{InSpireHEP} record.
For such manuscripts, identifiers are recorded 
to an alternative file (\texttt{APIQueryInputs.lostData[0]}).

\textbf{Phase 3 (Reprocessing Data)} 
serves as a bridge between fetching  data 
from online databases 
and creating plots. 
In  \texttt{doPhase3(\dots)},
the file containing bibliometric records 
(\texttt{APIQueryInputs.dataPath[0]}) 
is loaded (\texttt{pd.read\_csv(\dots)})
and data are re-record in new files
(\texttt{APIQueryInputs.plotData[N]}),
each retaining only the data necessary for specific plots. 

\textbf{Phase 4 (Plotting)} 
is the final phase of the script's default configuration.
In  \texttt{doPhase4(\dots)},
the data files prepared in \textbf{Phase 3} 
(\texttt{APIQueryInputs.plotData[N]}) are loaded
(\texttt{pd.read\_csv(\dots)})
and plotted as histograms.
Composite quantities (ratios and averages) 
are also computed and plotted.
Axes and labels are populated using
input information (\textbf{Phase 0}).
Finally, plots are saved
as \texttt{png} files 
(\texttt{APIQueryInputs.imgPaths[N]}).

\section{Pipeline Output}
\label{sec:examples}

We now show some results of the students' analysis.
Their investigation covers manuscripts submitted 
to the \textsc{arXiv} over the period  \confirm{1991 Jan.\ 01 -- 2025 Mar.\ 31}
\confirm{(34 years)} for six subject categories:
four in high energy physics and two in nuclear.
A manuscript's ``primary'' category must be one of these six.
Query inputs
are summarized in Table~\ref{tab:queryInputs}
and generated a raw dataset of 
\confirm{$N_{\rm arXiv}=329,035$} manuscripts.
Of these, \confirm{$N_{\rm tot}=328,936$} 
records were collected from \textsc{inSpireHEP}.
The ``missing'' \confirm{$\Delta N = 99$} records 
largely correspond to entries 
withdrawn\footnote{About half (a quarter)  
are from  the 2010s (2000s).}
from the \textsc{arXiv} and are not considered further.
In App.~\ref{app:extra},
 results for a one-year run
(\confirm{2022 Jan.~01--Dec.~31, 
with $N_{\rm tot}=12,099$ and $\Delta N = 2$}) 
are available.

Regarding computing resources, 
collecting, processing, and plotting 
\confirm{34 years} worth of papers required 
\confirm{about 3 days} of runtime. 
While \textbf{Phase 1} took \confirm{approximately half a day},
the bottleneck is \textbf{Phase 2}. Roughly,
our call rate to \textsc{InSpireHEP} is 
about \confirm{$r_{\rm call}\approx 2$ calls (1 call) per second
when connected to the internet by cable (Wi-Fi)}.
This translates to \confirm{$\Delta t_{\rm Phase\ 2} \approx N_{\rm tot} /  r_{\rm call} \sim 45\ (90)$ hours.}
We note that a poor internet 
connection can reduce the number of
found in \textbf{Phase 1}.
\textbf{Phases 3-4} took seconds.

By the end of \textbf{Phase 1}, 
about \confirm{240 MB} of memory was needed. 
\textbf{Phase 4} is the biggest memory consumer, 
requiring about \confirm{465 MB} to generate histograms.
(Due to the modular structure of the program, 
memory allocated for earlier phases is reused for later phases.)
Output files, including \texttt{png} files,
required about \confirm{10 MB of storage}.

\subsection{Submissions vs Day of Week}
\begin{figure}[t!]
\subfigure[]{\includegraphics[width=\columnwidth]{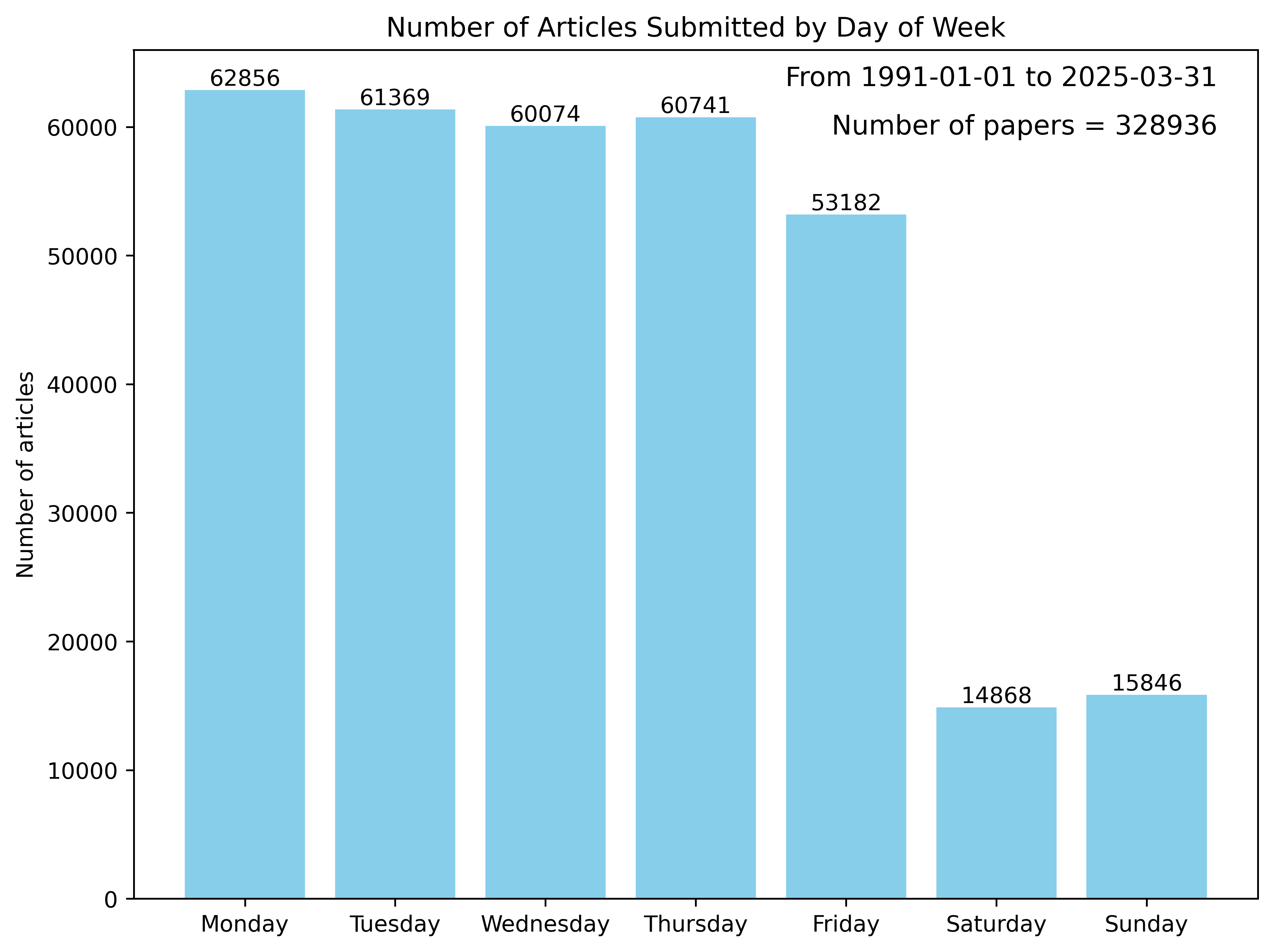}
\label{fig:bibAPI_submitted_by_weekday}
}
\subfigure[]{\includegraphics[width=\columnwidth]{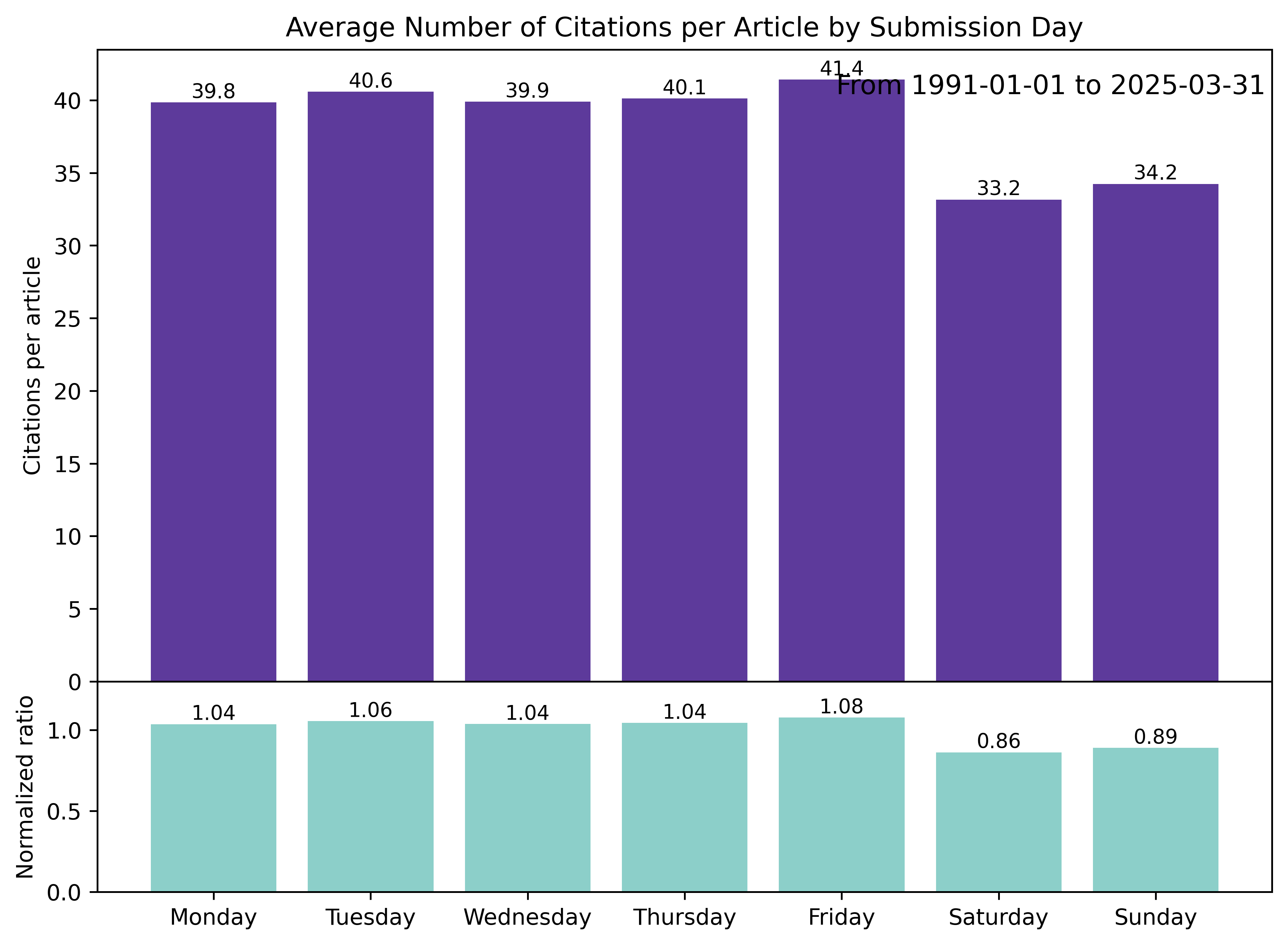}
\label{fig:bibAPI_citations_by_weekday}
}
  \caption{(a) 
  The total number of articles submitted to the \textsc{arXiv} as a function of day.
  (b) Upper: the average number of citations per article $\mathcal{C}_D$ as a function of day.
  Lower: Same as upper but normalized to the seven-day average 
  $\langle \mathcal{C}_7\rangle$.}
\label{fig:weekday}
\end{figure}

We start with Fig.~\ref{fig:bibAPI_submitted_by_weekday},
where we show as a baseline 
the total number of articles $(N_{\rm sub})$ submitted 
on each day of the week $(D)$, 
as defined by New York local time.
For each day, including weekends, 
\confirm{$\mathcal{O}(10^4)$} manuscripts were submitted to the \textsc{arXiv}.
This translates to an analyzing power (assuming counting statistics) of about 
\confirm{$\delta N = 1/\sqrt{N}\sim 0.4\%-0.8\%$}.
Interestingly,
while timezones can account for many Sunday submissions 
(Asia and Europe are ahead of New York), 
the volume of Saturday submissions suggests 
many weekend submissions.
Summing over the week recovers the full dataset:
\begin{align}
    \sum_{D\in\{\rm days\ of\ the\ week\}}\ N_{\rm sub}(D)\ =\ N_{\rm tot}\ 
    =\ \confirm{328,936}\ .
\end{align}

\begin{table}[t!]
\centering
\resizebox{\columnwidth}{!}{
\begin{tabular}{ c }
\hline  \hline
 Date Range: \confirm{1991 Jan.\ 01 -- 2025 Mar.\ 31} 
 \\
 \hline
HEP categories: experiment (\texttt{hep-ex}), theory (\texttt{hep-th}),\\ 
phenomenology (\texttt{hep-ph}), lattice (\texttt{hep-lat}) \\
\hline
Nucl categories: experiment (\texttt{nucl-ex}),
theory (\texttt{nucl-th})
 \\
 \hline \hline
\end{tabular}
}
\caption{Parameters for \textsc{arXiv} query in Sec.~\ref{sec:examples}.}
\label{tab:queryInputs}
\end{table}

In Fig.~\ref{fig:bibAPI_citations_by_weekday}
the upper panel shows the average number of citations per article $(\mathcal{C})$ 
as a function of submission day. 
The average $\mathcal{C}$ is defined as 
the sum of all citations $(c_i)$ over all articles $(i)$ submitted on a given day of the week
divided by the total number of submitted manuscripts on the same day. 
Symbolically, this ratio is given by
\begin{align}
    \mathcal{C}(D)\ =\ \frac{\sum_{i\in\{\rm articles\}}\ c_i(D)}{N_{\rm sub}(D)}\ .
\end{align}

At this level of granularity 
the average $\mathcal{C}$ 
appears uniform over the workday submissions
 but 
drops for weekend submissions.
We find the means 
and standard deviations 
for M-F $(\langle \mathcal{C}_5\rangle$, $\sigma_{5})$
and the full week 
$(\langle \mathcal{C}_7\rangle$, $\sigma_{7})$ are:
\begin{subequations}
\begin{align}
    \langle \mathcal{C}_5\rangle &= \sum_{D={\rm M}}^{\rm F} \mathcal{C}(D)/5\ 
    \approx \confirm{40.4}\ ,
    \\
    \frac{\sigma_{5}}{\langle \mathcal{C}_5\rangle} &= 
    \frac{1}{\langle \mathcal{C}_5\rangle}
    \sqrt{\frac{1}{5}\sum_{D={\rm M}-{\rm F}}\left(\mathcal{C}(D)-\langle \mathcal{C}_5\rangle\right)^2}\ \approx\ 
    \confirm{1\%}\ ,
    \\
    \langle \mathcal{C}_7\rangle &
    \approx \confirm{38.5}\ ,\quad  
    \frac{\sigma_{7}}{\langle \mathcal{C}_7\rangle} \approx\ 
    \confirm{8\%}\ .
\end{align}
\end{subequations}
In words, the average \textsc{arXiv} submission in HEP and nuclear physics 
has about \confirm{$38.5\pm 8\%$ citations}
regardless of the day it is submitted
to the database,
but about 
\confirm{$40.4\pm 1\%$ citations}
if submitted during the weekday.
We further illustrate the 
(non)uniformity of citation  averages 
in the lower panel of Fig.~\ref{fig:bibAPI_citations_by_weekday},
which shows the upper panel normalized by the 7-day mean, 
i.e., the ratio $\mathcal{C}(D) / \langle\mathcal{C}_7\rangle$.

Fig.~\ref{fig:bibAPI_citations_by_weekday} 
reveals an interesting finding.
When a manuscript is submitted to the \textsc{arXiv},
its appearance is delayed by moderators who screen for 
plagiarism, etc. 
Wednesday submissions, for example, usually appear Friday.
Since there are no \textsc{arXiv} updates over the weekend,
there is a myth 
in the HEP theory community 
that articles submitted on Wednesdays 
are better cited 
because they have more exposure over the weekend.
\confirm{Our analysis does not support this claim:
Wednesday submissions (Friday appearance)
generate an average number of citations.}
However, data do suggest 
that weekend submissions garner 
about $6-7$ fewer citations, 
which is over  $10\times\sigma_5$
below from the weekday submission 
$\langle \mathcal{C}_5\rangle$.

\begin{figure}[t!]
\includegraphics[width=\columnwidth]{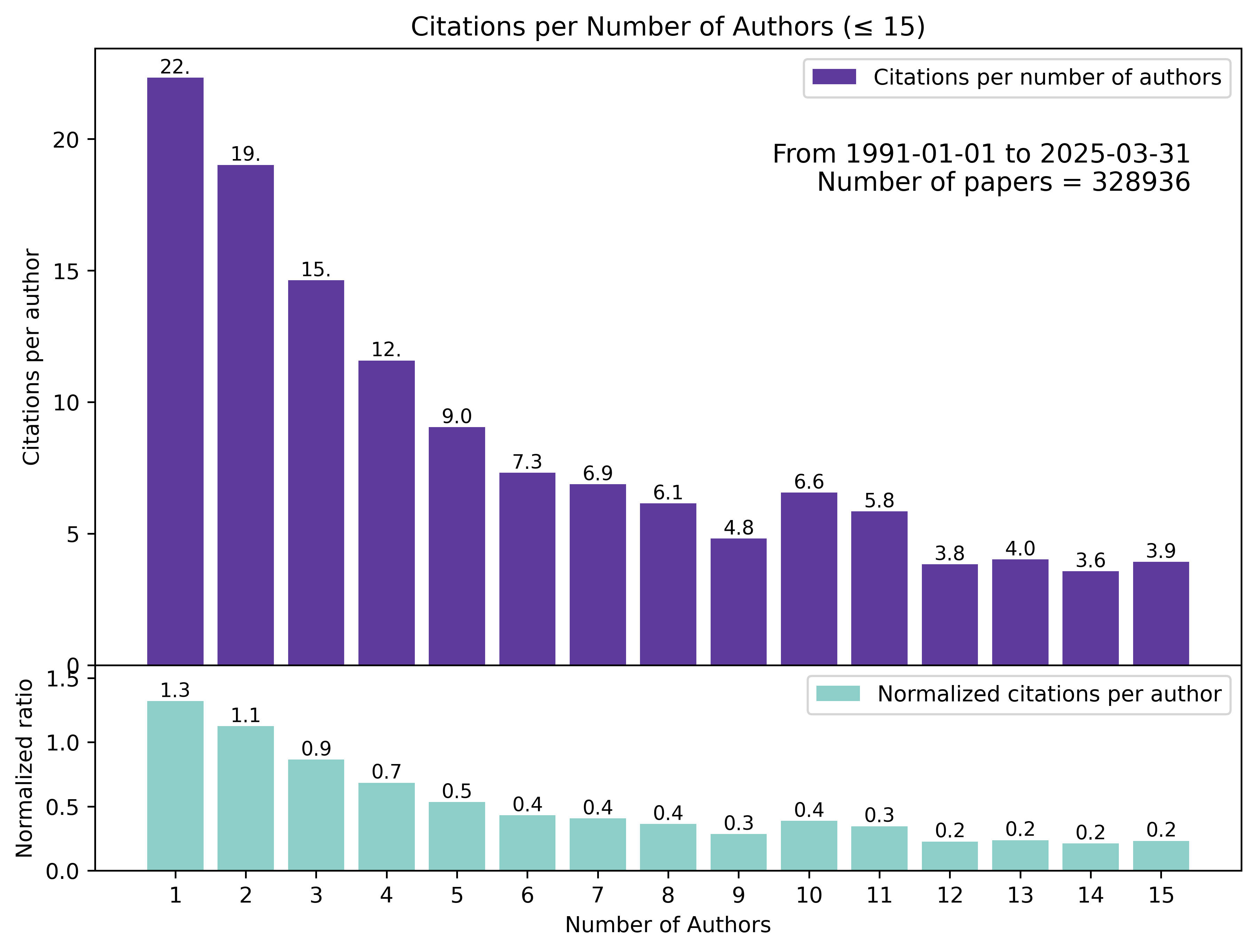}
  \caption{Upper: The average number of citations per author $\mathcal{C}(n_{\rm auth})$
  as a function of a manuscript's authorship size.
  Lower: Same as upper but normalized to the mean $\langle \mathcal{C}_{15}\rangle$.}
\label{fig:bibAPI_citations_by_authors}
\end{figure}

\subsection{Citations vs Authorship}

In Fig.~\ref{fig:bibAPI_citations_by_authors},
the upper plot shows as a function of 
a manuscript's authorship size $(n_{\rm auth})$
the average number of citations per author $\mathcal{C}(n_{\rm auth})$.
Using the relationship 
\begin{align}
    \mathcal{C}(n)\ = 
    \frac{\sum_{i\in\{\rm articles\}} 
    c_i(n)/n}{N_{\rm sub}(n)} 
    =
    \frac{\sum_{i\in\{\rm articles\}} c_i(n)}{n\ N_{\rm sub}(n)}\ ,
\end{align}
the average $\mathcal{C}(n)$ is equal to 
the sum of all citations $(c_i)$ over all articles $(i)$ 
with a given authorship size $(n)$,
divided by the authorship size and 
the number of submitted articles $(N_{\rm sub})$ with the same number of authors.
To further illustrate overall trends, 
we show in the lower plot the same data as in the upper plot 
normalized by the mean  
over all submitted papers
$\langle \mathcal{A}_{15}\rangle$ 
with $n_{\rm auth}\leq15$, 
\begin{align}
    \langle \mathcal{A}_{15}\rangle 
    &=
    \frac{\sum_{i\in\{\rm articles\}} 
    c_i(n)/n}{N_{\rm tot}} 
    \\
    &=
    \frac{1}{N_{\rm tot}}
    \sum_{n=1}^{15} 
    \mathcal{C}(n)\ N_{\rm sub}(n)
    \approx \confirm{17.0}\ .
\end{align}

We find that the average number of citations per author  
\confirm{decreases almost uniformly with increasing collaboration size}.
The exception is for \confirm{$n_{\rm auth}=10-11$ authors.
Here, 
the distribution reaches a local maximum with $\mathcal{C}(n_{\rm auth}=10-11)\approx6.6-5.8$} 
citations per author,
which translates to about 
\confirm{$64-66$} total paper citations.
For reference, \confirm{$\mathcal{C}(n_{\rm auth}=5,\ 9,\ 12)\approx 9.0,\ 4.8,\ 3.8$},
and translate to about $45-46$ total 
paper citations on average.
However, we caution making 
too strong an inference.
For manuscripts with $n=1-3$,
there are about 
$N_{sub}(n=1-3)\approx7.1-8.4\times10^4$
entries.
For manuscripts with  $n\geq10$ authors,
there are fewer than 
$3100$ entries.

\subsection{Citations vs Subject Category}

\begin{figure}[t!]
\subfigure[]{\includegraphics[width=\columnwidth]{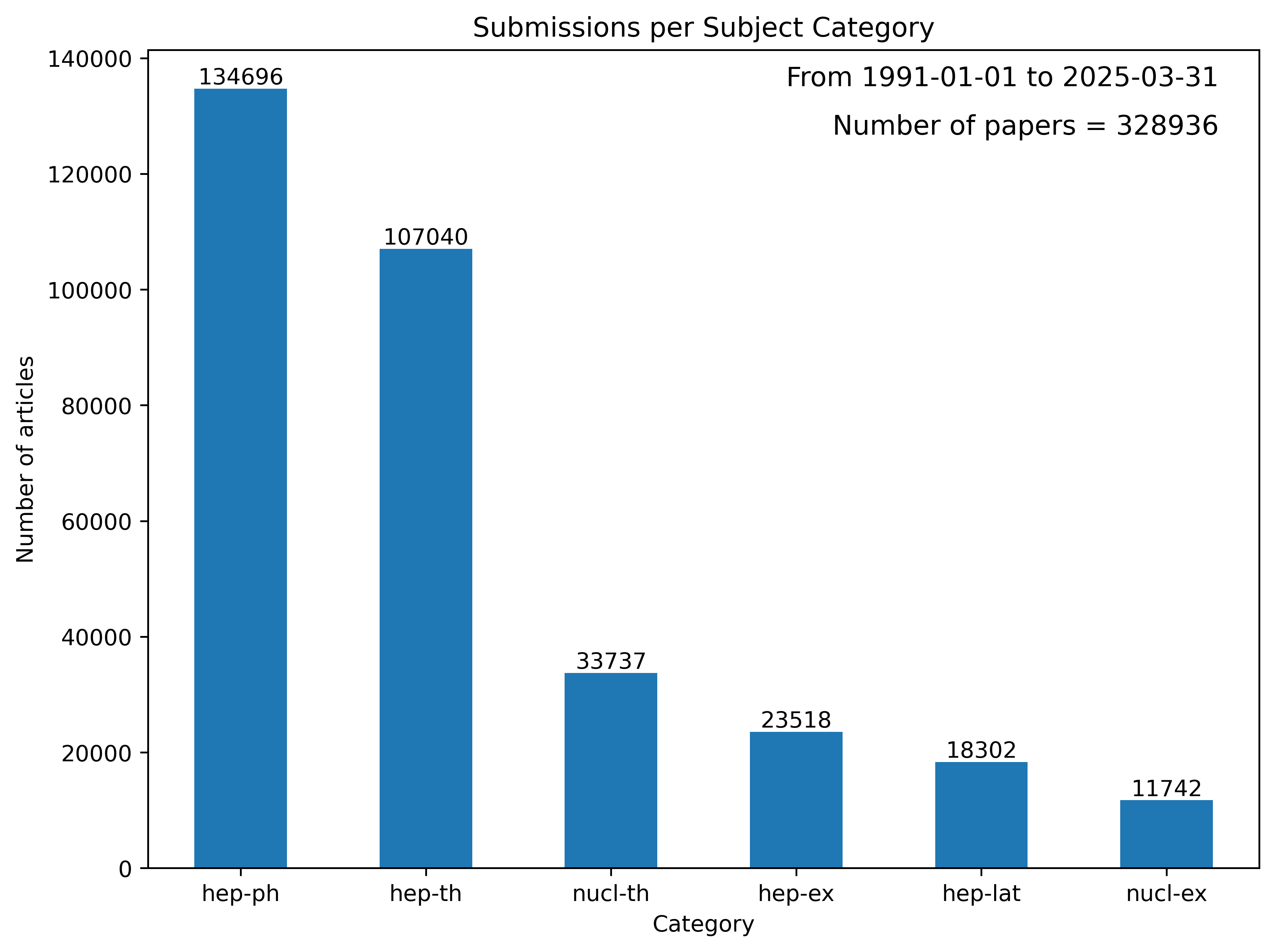}
\label{fig:bibAPI_submitted_by_subject}}
\subfigure[]{\includegraphics[width=\columnwidth]{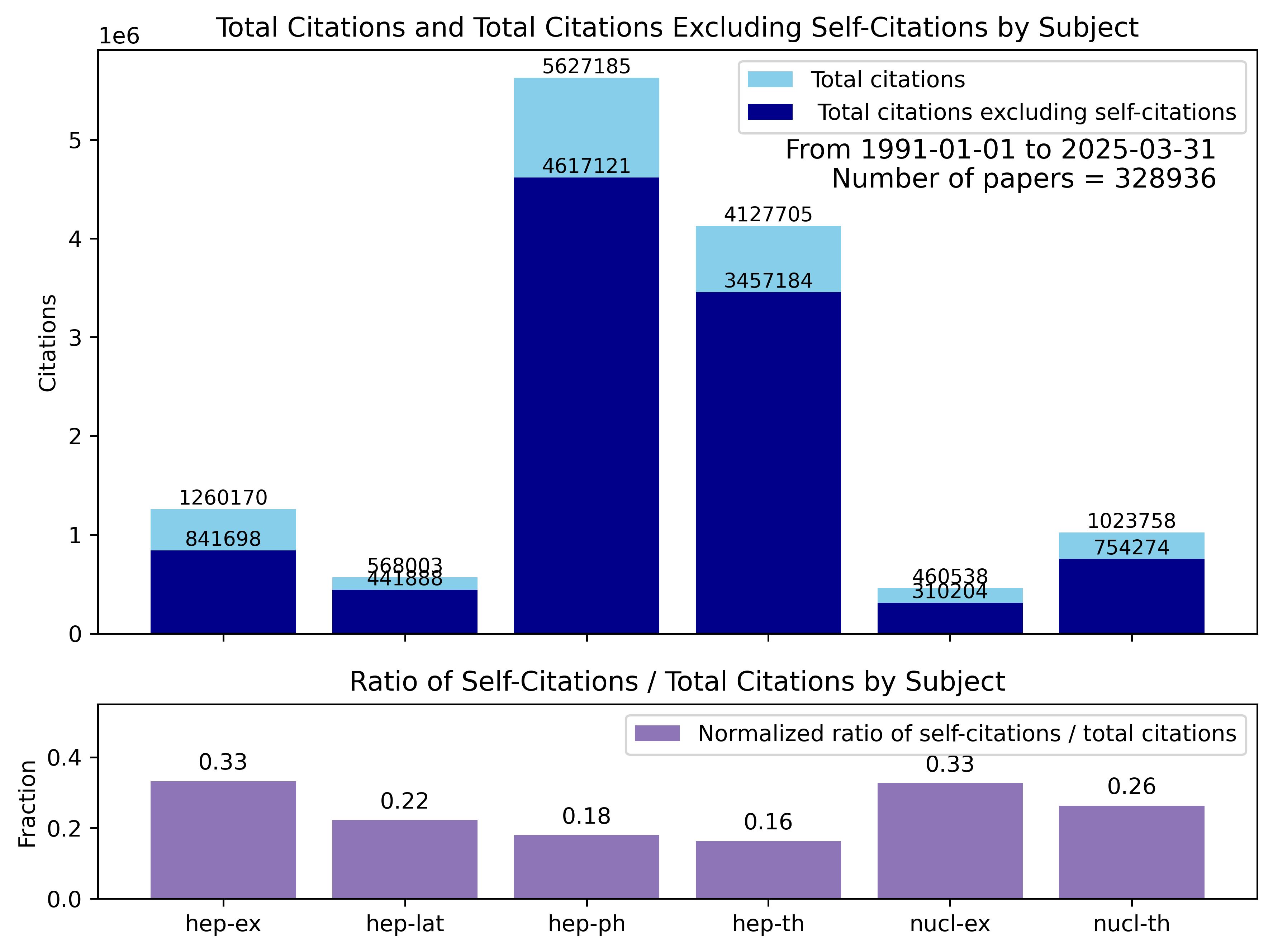}
\label{fig:bibAPI_citations_by_subject}}
  \caption{(a) Total papers submitted by subject category.
  (b) Upper: Total number of citations of all manuscripts by subject,
  including (light) and excluding (dark shading) self-citations.
  Lower: Self-citations as a fraction of total.}
\label{fig:bibAPI_subject}
\end{figure}

Finally, in Fig.~\ref{fig:bibAPI_subject} we show the dependence 
on individual subject categories.
As a baseline we show in Fig.~\ref{fig:bibAPI_submitted_by_subject}
the total number of papers submitted by category.
The differences in paper count highlight the different norms 
and expectations within 
particle and nuclear physics.

In the upper panel of 
Fig.~\ref{fig:bibAPI_citations_by_subject} we show as a function of 
subject category 
the total number of citations 
of all papers within that category
(light shading).
Here, total citations 
$c_{\rm tot}({\rm sub})$
within a category 
is given by the sum of 
individual paper 
citations $c_i({\rm sub})$;
this  
includes  
self-citations\footnote{Defined by \textsc{InSpireHEP} as ``citations from the same collaboration or any of the authors of the paper being cited.'' See  \href{https://help.inspirehep.net/knowledge-base/citation-metrics/}{help.inspirehep.net/knowledge-base/citation-metrics/}.}  
$c_i^{\rm self}({\rm sub})$
and citations by other authors
$c_i^{\rm others}({\rm sub})$.
Symbolically, these counts are related by
\begin{subequations}
\begin{align}
c_{\rm tot}({\rm sub})\
 &=\
 c_{\rm tot}^{\rm self}({\rm sub})\
 +\
 c_{\rm tot}^{\rm others}({\rm sub})\ ,
 \\
 c_{\rm tot}^{\rm self\ (others)}({\rm sub}) &=\ \sum_{i\in\{\rm articles\}} c_i^{\rm self\ (others)}({\rm sub})\ .
\end{align}
\end{subequations}

Also shown in Fig.~\ref{fig:bibAPI_citations_by_subject}
as a function of subject 
are the total number 
of citations excluding 
self-citations (dark shading).
In the lower panel, we show 
the fraction of 
self-citations $(\mathcal{S})$, 
which can be written as the ratio
\begin{align}
   \mathcal{S}({\rm sub})\ =\
   \frac{c_{\rm tot}({\rm sub}) - c_{\rm tot}^{\rm others}({\rm sub})}
   {c_{\rm tot}}\ .
\end{align}

We observe that experimental 
papers self-cite at a slightly
higher rate than 
theoretical works,
with  
self-citations constituting 
$\mathcal{S}\sim 33\%$ 
of all citations in experimental works
and 
about 
$\mathcal{S}\sim 20\%$
of all citations in theoretical works.
We avoid commentary on practice of  
self-citing but note that our dataset 
includes conference proceedings 
submitted to the \textsc{arXiv}.
Experimental collaborations 
frequently produce 
conference proceedings that
often reference multiple 
experimental papers.

\section{Applications and Extensions}
\label{sec:extensions}

For students with more advanced computing skills,  
the ``farm-to-table'' data analysis pipeline of  
Fig.~\ref{fig:apiFlow_arXivInSpire} 
can be applied and extended in several ways.
For instance: during \textbf{Phase 4}, 
one can assume a statistical model 
(such as those available in \texttt{scipy}) and fit the data.
One can train a machine learning algorithm 
(e.g., regression models or neural networks as available in \texttt{sklearn})
to explore more sophisticated correlations.
Beyond statistical modeling, 
one can also explore computational security protocols.

In large-scale science experiments, security is an important but seldom-discussed topic.
Datasets need to be securely stored but also accessible by credible users.
To access non-public datasets of 
experiments and observatories,
one generally uses login credentials.
A well-known instance of an analysis pipeline
employing login credentials 
is the \textit{online} monitoring system of an experiment.

The Large Hadron Collider and its 
experiments, for example, 
share an automated data acquisition (DAQ) 
monitoring system 
that gives online status reports
in real time\footnote{Publically available at \href{https://op-webtools.web.cern.ch/vistar/}{op-webtools.web.cern.ch/vistar/}.}.
The monitor's pipeline follows Fig.~\ref{fig:apiFlow_arXivInSpire}:
For each system, 
diagnostic data  are 
 collected (\textbf{Phases 1--2}), 
processed (\textbf{Phase 3}), 
and plotted (\textbf{Phase 4}) on a local server.
Plots are then copied to a web server (\textbf{Phase 5}).
Another example is the 
SuperNova Early Warning System\footnote{Details available at 
\href{https://snews2.org/}{https://snews2.org/}},
where a supernova candidate in one telescope/observatory 
transmits an alert to others  through 
a credentialed, one-way ``query'' (\textbf{Phases  5}).
For such interfacing between known severs,
\texttt{scp}/\texttt{ssh} commands with RSA keys are used.
For interfacing between unknown servers (via APIs), 
``access tokens'' are a popular alternative.

To this end, we further explored whether 
API security protocols could be added to 
 Fig.~\ref{fig:apiFlow_arXivInSpire},
and therefore also into coursework.
Again, the outcome was positive.
In this case, an early career postdoc 
built an ``export pipe'' that 
posts the plots from \textbf{Phase 4} 
 to social media accounts.
We now describe \textbf{Phase 5} of the \texttt{BibAPI} template.

\subsection{Phase 5: Exporting to Social Media}
\label{sec:extensions_socials}

\textbf{Phase 5 (Exporting)} of \texttt{BibAPI} 
exports plots from \textbf{Phase 4}
to social media accounts via their APIs.
We focus on the platforms
\textit{Bluesky} and \textit{Twitter/X}
to highlight the similar procedures among media platforms,
including \textit{Facebook} and \textit{LinkedIn}.
Credentialed APIs operate like public APIs (e.g., \textsc{InSpireHEP}).
The main difference is that a ``client'' calls URLs/URIs on behalf of a user.

In practice, instead of directly calling a URL in a script (e.g., with \texttt{urlopen(\dots)}),
a client must first be initialized, typically with calls of the 
form\footnote{API software libraries are  
maintained by the platforms themselves.
Usually, these are free to encourage 
the development and commercialization of 
API applications, i.e., ``web apps.''} 
\texttt{client = ClientClass(\dots)} and \texttt{client.login(\dots)}. 
The client then calls a \textit{URI}, i.e.,
an elongated URL containing encrypted credentials {and} content (markup text),
through a ``post'' command\footnote{Using a ``get'' command of the form \texttt{client.get(\dots)} will instead retrieve data from the platform as in \textbf{Phases 1} and \textbf{2}.} of the form \texttt{client.post(\dots)}.
The platform interprets the URI and carries out the embedded instructions.
To post media (e.g., \texttt{png} files), 
one may need to first upload the file via an upload call of the form \texttt{client.upload(\dots)}.
This type of call returns the URL of the file, which can be passed to  
\texttt{client.post(\dots)}.

In  \texttt{doPhase5(\dots)}, the above procedure is carried out.
API clients for 
\textit{Bluesky} (\texttt{atproto.Client(\dots)}) and
\textit{Twitter/X} (\texttt{tweepy.Client(\dots)})  
are first initialized.
Figures from \textbf{Phase 4} (\texttt{APIQueryInputs.imgPaths[N]}) are then uploaded
and published 
on the platforms
(\texttt{send\_images(\dots)} and \texttt{create\_tweet(\dots)}).

In the \texttt{client} calls above, ``tokens'' are passed as arguments. 
These are pseudo-login credentials that anonymize real credentials.
In principle, only the platforms can map ``tokens'' to real credentials.
Tokens can be fed into URIs, given a fixed lifetime,
and granted restricted account permissions, 
e.g., \textit{only} read.
Unlike real credentials, 
a single set of login tokens can be shared.

To generate API tokens, 
a developer's account on the platform must be created 
and tied to an existing social media account.
Importantly, free developer accounts 
can be obtained for academic/educational use.
Under such licensing agreements, 
 API calls from 
 pipelines-turned-web-apps 
\textit{cannot} be commercialized. 
Under alternative licenses, 
implementations of 
Fig.~\ref{fig:apiFlow_arXivInSpire}
\textit{can} be commercialized.
This exemplifies  how Big Data methods used in physics \& astronomy research
are applicable in industry.

\subsection{An Automated Online DAQ Monitor}
\label{sec:examples_daq}

With \textbf{Phase 5}, 
the 
pipeline depicted in Fig.~\ref{fig:apiFlow_arXivInSpire}
mirrors the Large Hadron Collider's online DAQ monitor.
However, unlike \texttt{BibAPI}, 
real online DAQ monitors are automated and run continuously.
Turning \texttt{BibAPI} into a ``bot'' is actually straightforward.
One simply needs a job scheduler, 
such as \texttt{cron} on \textit{Linux} systems. 

\begin{figure}[t!]
  \includegraphics[width=\columnwidth]{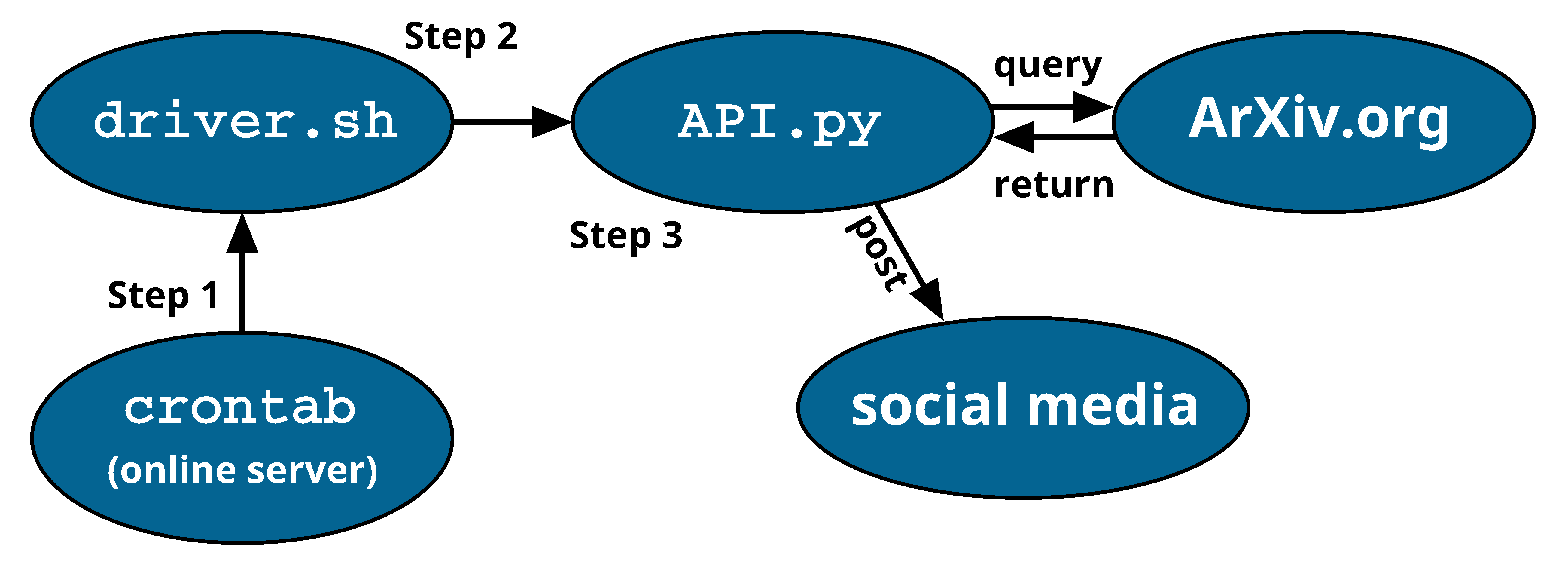}
  \caption{Illustration of an automated API/social media bot.}
\label{fig:apiFlow_SocialMediaBot}
\end{figure}

As depicted in Fig.~\ref{fig:apiFlow_SocialMediaBot},
a two-to-three-line \texttt{crontab} operating daily 
(for \textsc{arXiv} queries) 
or every minute (for DAQ monitoring) 
on an online server can call a two-to-three 
line shell script (\texttt{driver.sh})
that subsequently calls an API script (\texttt{API.py}).
The script can then make the appropriate queries and 
posts to online platforms.
In the context of science communication and outreach,
we also report the successful launch of a bot 
that checks the \textsc{arXiv} daily for 
new research papers by certain research networks 
and advertises the work on social media\footnote{Details are available at \href{https://gitlab.cern.ch/riruiz/public-projects/-/tree/master/COMETA_SocialsAPI}{gitlab.cern.ch/riruiz/public-projects/-/tree/master/COMETA\_SocialsAPI}.}.

\section{Discussion and Conclusions}
\label{sec:conclusion}

Many disciplines in the physical sciences,
including 
astrophysics, nuclear physics, and particle physics
have become synonymous with Big Data.
Among other traits,
Big Data is
characterized
by the \textit{fast} handling of large datasets
(in excess of $\mathcal{O}(10^6)$ entries),
 strict memory and storage management,
 and
interfacing directly with databases,
possibly with security protocols.
In other words: advanced computing skills.
It is likely that these skills will 
be increasingly integrated into 
physics \& astronomy curricula,
particularly at the undergraduate level~\cite{aaptGuestEd}.
However, the fact is that 
physics \& astronomy
students face barriers
to learning these skills~\cite{ShahKnaub:2024}.

In this work, we have presented a
``farm-to-table'' analysis pipeline
that collects, processes, and plots
(biblometric) data from
two public databases
that are well-known in the physics community.
The pipeline (Sec.~\ref{sec:method})
was inspired by those used in 
contemporary physics \& astronomy research
and can be implemented (Sec.~\ref{sec:implementation})
using 
\textit{Python} libraries
typically introduced in undergraduate computing courses.
Our overarching motivation 
is to demonstrate that advanced data analysis skills
can successfully be introduced at the undergraduate level.
These are skills that are typically taught at the
PhD level but  often are needed
by students earlier in their training.

The key to the pipeline's operation and success, 
is the use of Application Program Interfaces,
which are prevalent in industry, the public sector, and Big Science.

To illustrate the classroom potential of the pipeline,
we showed in Sec.~\ref{sec:examples}
a simplified analysis of bibliometric data 
from about \confirm{330k}
manuscripts in high energy and nuclear physics 
submitted to the \textsc{arXiv} over a \confirm{34-year period}.
The analysis itself was authored by two undergraduate students 
during a 4-week internship and is publicly available.
In Sec.~\ref{sec:extensions}, 
potential applications and extensions of the pipeline 
are discussed, 
including security protocols and pathways commercialization.
With suitable tailoring,
we believe that some or all parts of the pipeline
can be incorporated into
undergraduate curricula,
including at the first- and second-year level.

\section*{Acknowledgments}
The authors thank Hal Haggard and Alexis Knaub for discussions.
The authors also thank the  \textsc{arXiv} and \textsc{InSpireHEP} for 
the use of its open access interoperability.
This work was partially funded by the Polish Academy of Sciences under the
`Funding for the dissemination and promotion of scientific activities' 
program (PAN.BFB.S.BUN.442.022.2024).
SD and RR acknowledge the support of Narodowe Centrum Nauki under Grants 
2019/34/E/ST2/00186 and 
2023/49/B/ ST2/ 04330 (SNAIL). 
The authors acknowledge support from the COMETA COST Action CA22130.


\bibliography{bibAPI_refs.bib}

\appendix

\section{Computational Setup}
\label{app:setup}

The \texttt{BibAPI} template was developed and tested
on \textit{Linux} and \textit{MacOS} operating systems,
but not \textit{Windows}. 
However the use of these template on \textit{Windows} 
is not expected to be problematic. 
The script does not require significant computing power, 
as such any  basic, single-core, 
off-the-shell personal laptop or desktop computer 
with a stable internet connection should have sufficient 
memory and disk resources to run it.

\texttt{Python3} was used (more specifically \texttt{Python3.12} and \texttt{Python3.13}) with several libraries and modules. To make requests to the databases APIs, \texttt{urllib} was used alongside \texttt{feedparser} to parse the data obtained from the queries. The raw data saved and stored in external \texttt{txt} files are loaded using the \texttt{pandas} library, which is widely used in Data Science.
Pandas facilitates easy manipulation of data frames / data sets. 
Plots were produced using the commonly used \texttt{matplotlib} library. 

The social media APIs from  Sec. \ref{sec:extensions_socials} 
require  the 
\texttt{tweepy} (\textit{Twitter/X}) and \texttt{atproto} (\textit{Bluesky}) modules. It is possible to extend \textbf{Phase 5} to \textit{Facebook} using the \texttt{facebook-sdk} module.  
Listing 1 present a snippet of the code used to authenticate 
a client on \textit{Twitter/X} and post to social media.
Posting to other social media platforms follows a similar procedure.
By default, \textbf{Phase 5} is skipped in the 
\texttt{BibAPI} template and associated modules are \textit{not}
needed to run \textbf{Phases 1-4}.

A summary of specialized  libraries used, 
and the phases in which they are used, 
alongside the command lines to install them locally 
can be found in Tab. \ref{tab:pythonlibraries}.

\begin{table}[t!]
\centering
\begin{tabular}{ |c |c| c| l |}
\hline
\texttt{PY3} Module name & Versions & \textbf{Phase} & Local installation  \\
 \hline
 \texttt{urllib3} & 2.3.0/2.4.0 & 1,2,3,4 &
 \texttt{\$ pip3 urllib}\\
 \texttt{feedparser} & 6.0.11  & 1,2,3,4&
 \texttt{\$ pip3 feedparser}\\
 \texttt{pandas} & 2.2.3 & 1,2 &
 \texttt{\$ pip3 pandas}\\
 \texttt{matplotlib} & 3.8.3/3.10.3 & 4&
 \texttt{\$ pip3 matplotlib}\\
 \texttt{tweepy} & 4.15.0 & 5&
 \texttt{\$ pip3 tweepy} \\
  \texttt{atproto} & 0.0.61  & 5&
 \texttt{\$ pip3 atproto} \\
  \texttt{facebook} & 3.1.0 & 5&
 \texttt{\$ pip3 facebook-sdk} \\
 \hline
\end{tabular}
\caption{Specialized \textit{Python} libraries used in the project}
\label{tab:pythonlibraries}
\end{table}

\begin{lstlisting}[language=Python,caption=Creation of the OAuth1.0 and 2.0 clients and tweet publication with media]

import tweepy

# define OAuth1.0 client (for making post)
client = tweepy.Client(
    consumer_key=CONSUMER_KEY,
    consumer_secret=CONSUMER_SECRET,
    access_token=ACCESS_TOKEN,
    access_token_secret=ACCESS_TOKEN_SECRET
)

# define OAuth2.0 client (for uploading media)
apiKeys=tweepy.OAuthHandler(
kingdom.CONSUMER_KEY,
kingdom.CONSUMER_SECRET)

apiKeys.set_access_token(
kingdom.ACCESS_TOKEN,
kingdom.ACCESS_TOKEN_SECRET)

# upload media to platform
clientAPI=tweepy.API(apiKeys)
media = clientAPI.media_upload(pathtomedia)
meID = media.media_id_string

# simple test tweet
tweet="twitter API media test!"

# push post to platform
client.create_tweet(text=tweet,media_ids=meID)
\end{lstlisting}

\section{Usage and Options}
\label{app:usage}

After the libraries listed in Tab.~\ref{tab:pythonlibraries}
been installed, \texttt{bibAPI.py} 
can run from a terminal with the command
\begin{verbatim}
$ python ./bibAPI.py START END CAT1 CAT2 ...
\end{verbatim}
Here, \texttt{START} and \texttt{END}
are the start and end dates 
in \texttt{YYYY-MM-DD} format
of the \textsc{arXiv} query. 
\texttt{CATN} are subject categories 
as abbreviated by the \textsc{arXiv}.
For systems where \texttt{python} refers to \texttt{python2},
one should call \texttt{python3}.

\begin{figure}[t!]
\subfigure[]{\includegraphics[width=\columnwidth]{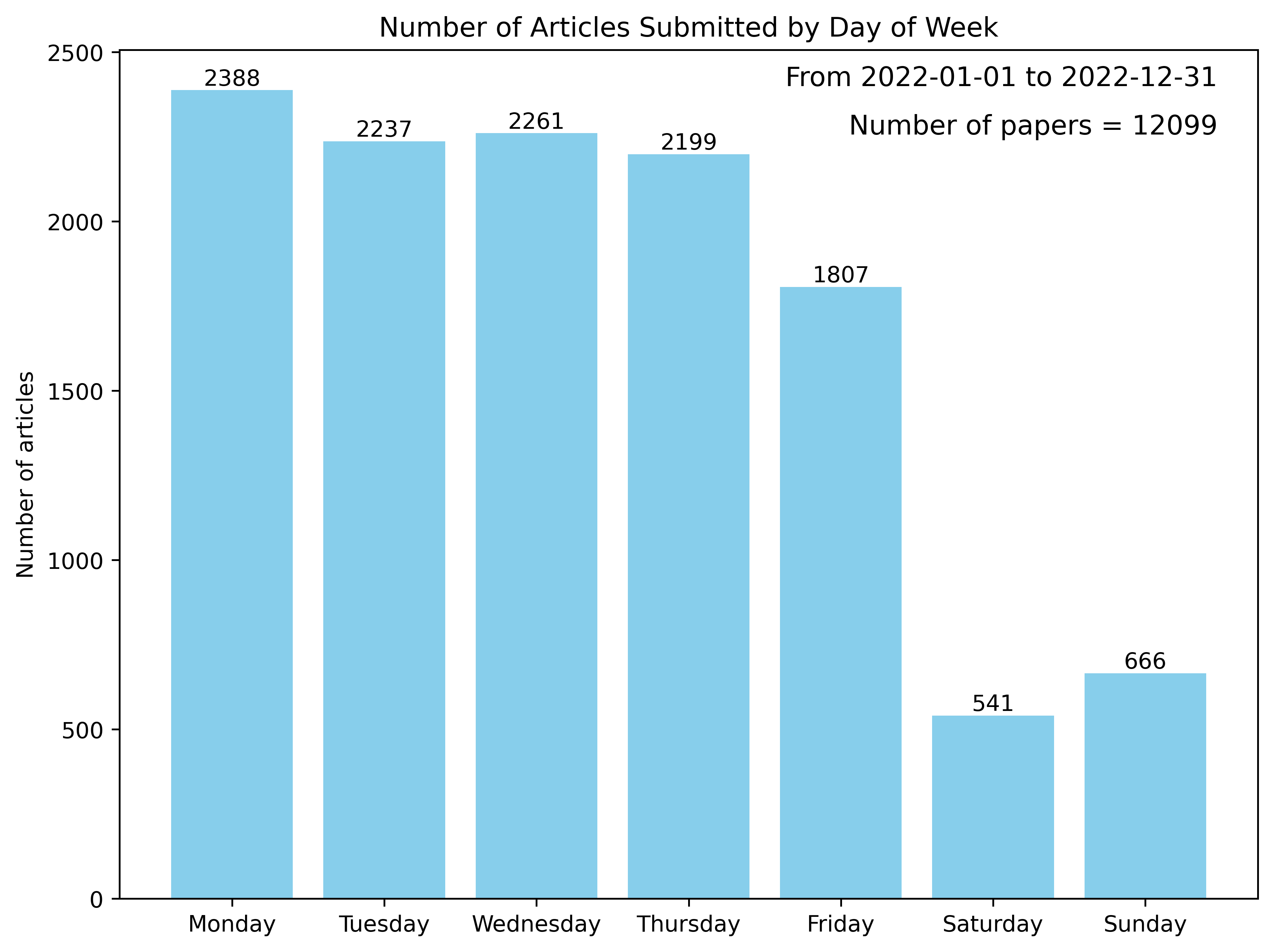}
\label{fig:bibAPI_submitted_by_weekday_2022}
}
\subfigure[]{\includegraphics[width=\columnwidth]{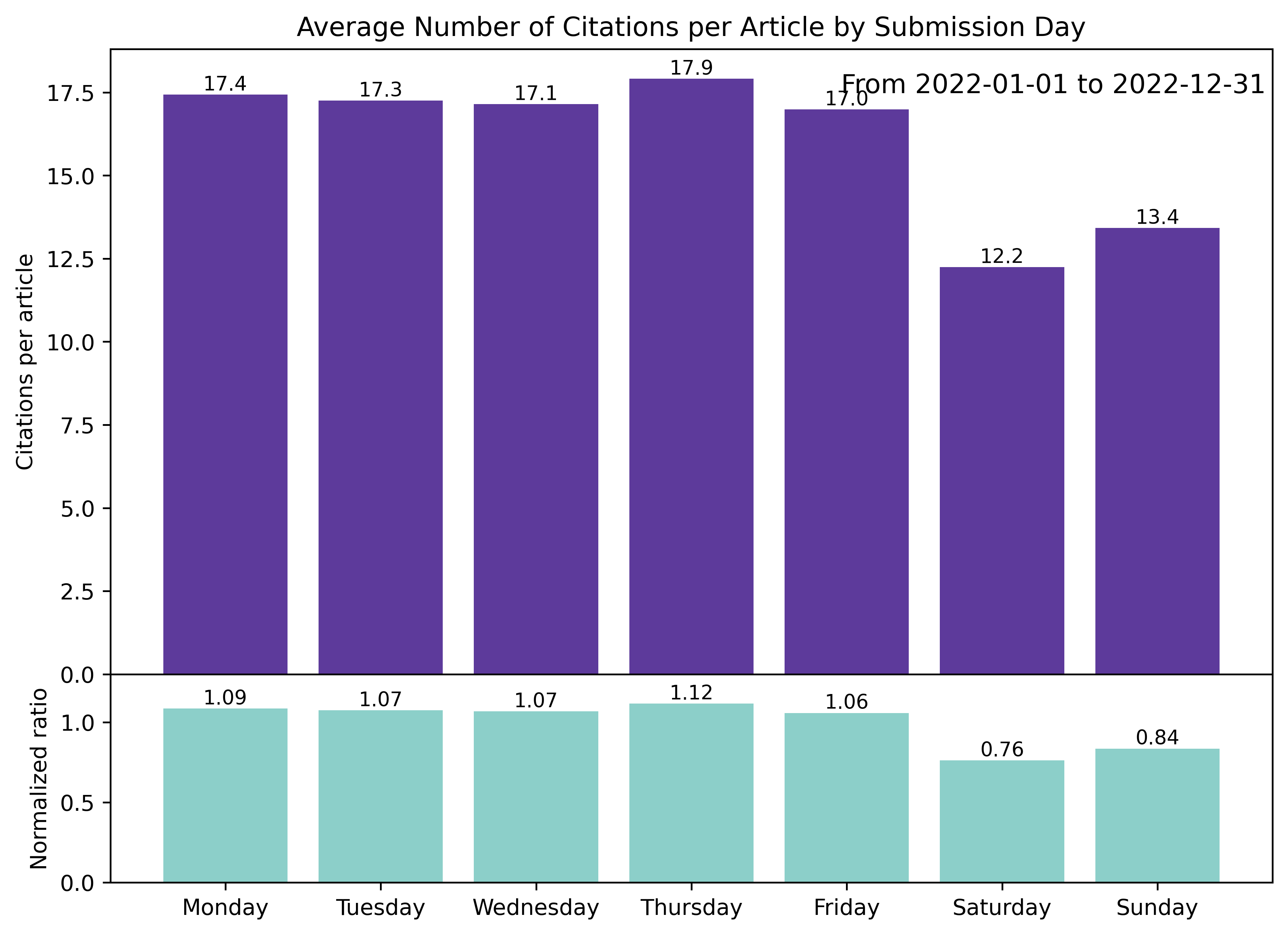}
\label{fig:bibAPI_citations_by_weekday_2022}
}
  \caption{(a,b) same as Fig.~\ref{fig:bibAPI_submitted_by_weekday}
  and
  Fig.~\ref{fig:bibAPI_citations_by_weekday}
  but for the date range 2022 Jan.~01 -- 2022 Dec.~31.}
\label{fig:weekday_2022}
\end{figure}

\begin{figure}[t!]
\includegraphics[width=.9\columnwidth]{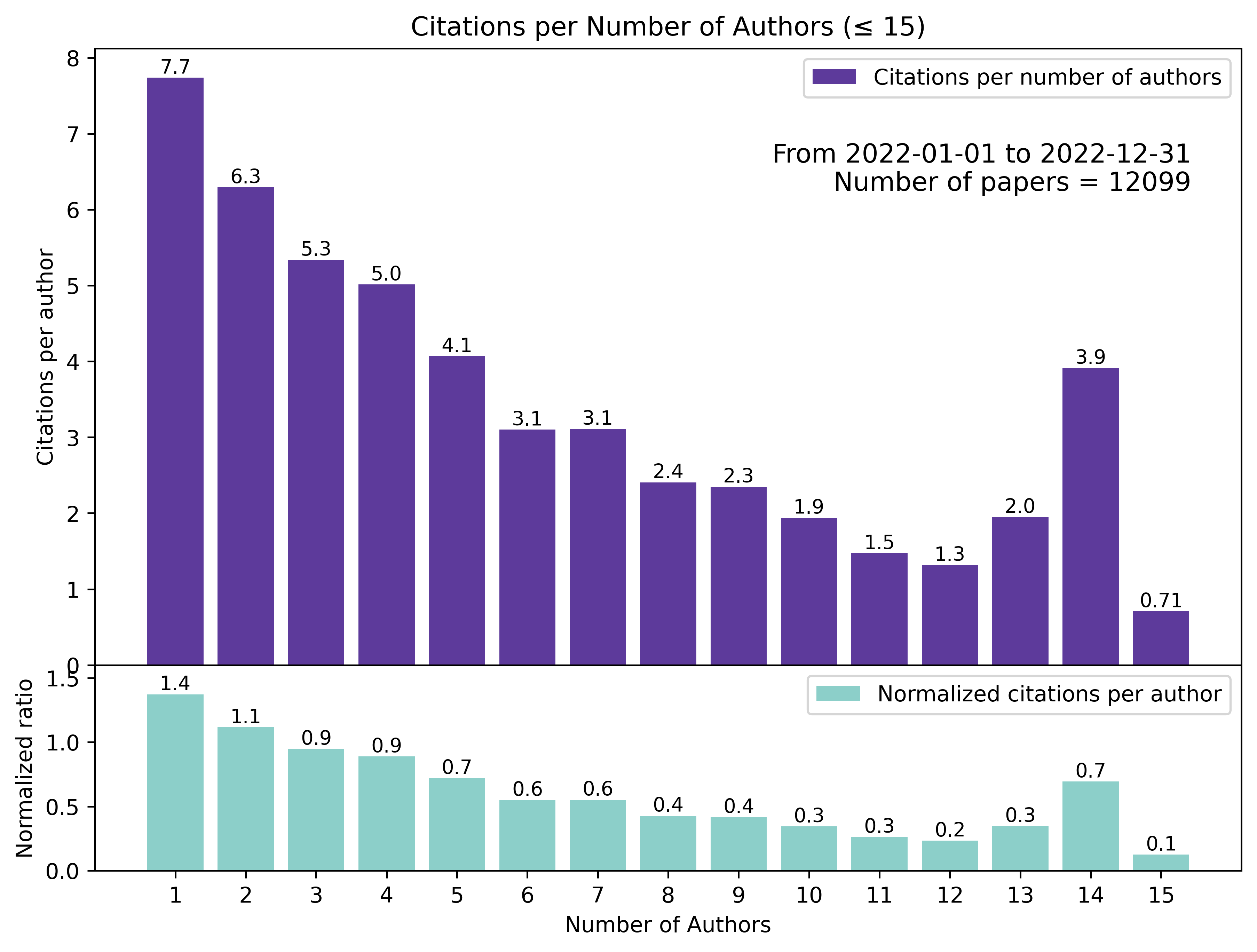}
  \caption{Same as 
  Fig.~\ref{fig:bibAPI_citations_by_authors}
  but for the date range 2022 Jan.~01 -- 2022 Dec.~31.}
\label{fig:bibAPI_citations_by_authors_2022}
\end{figure}

\begin{figure}[t!]
\subfigure[]{\includegraphics[width=.9\columnwidth]{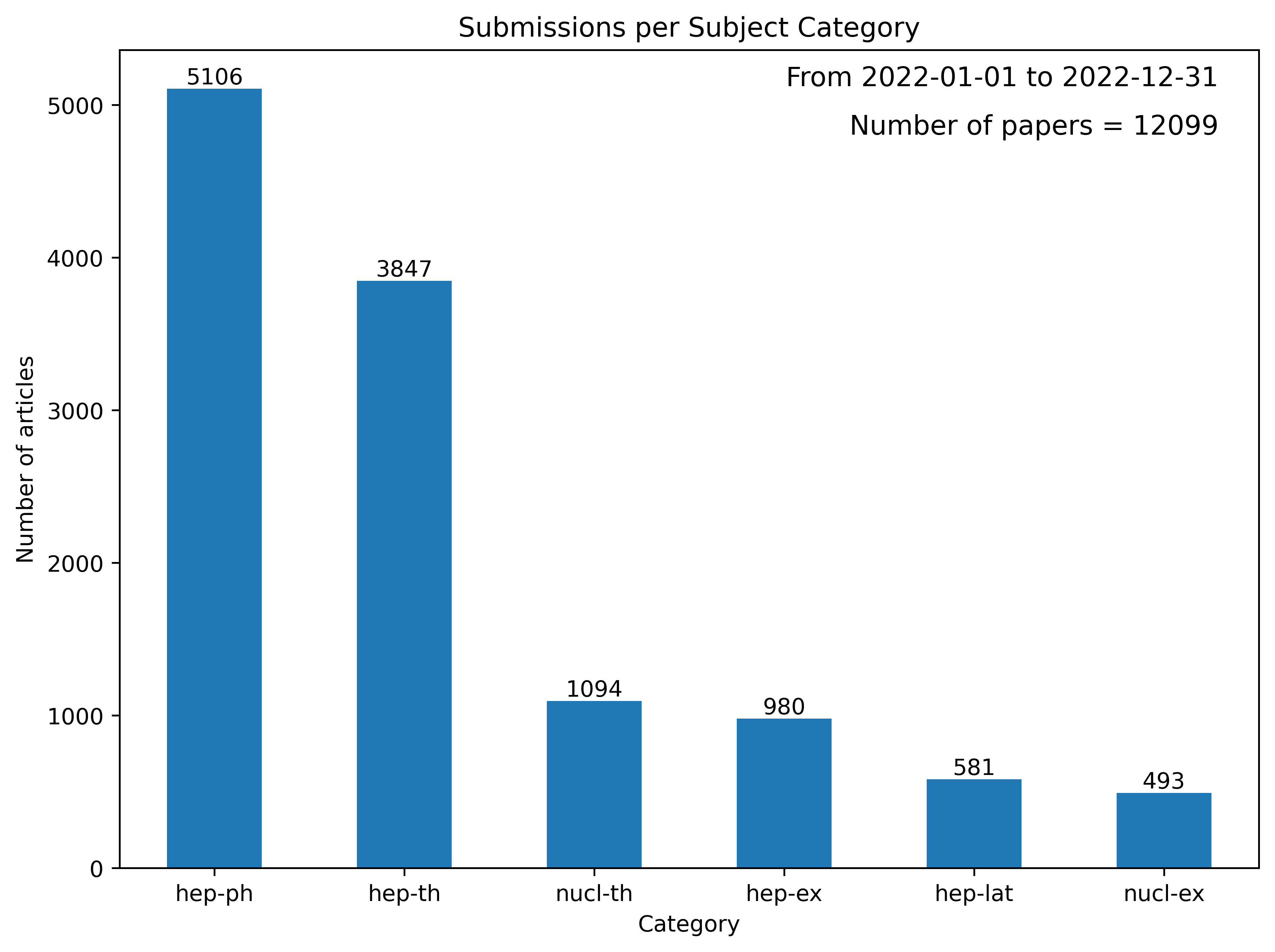}
\label{fig:bibAPI_submitted_by_subject_2022}
}
\subfigure[]{\includegraphics[width=.9\columnwidth]{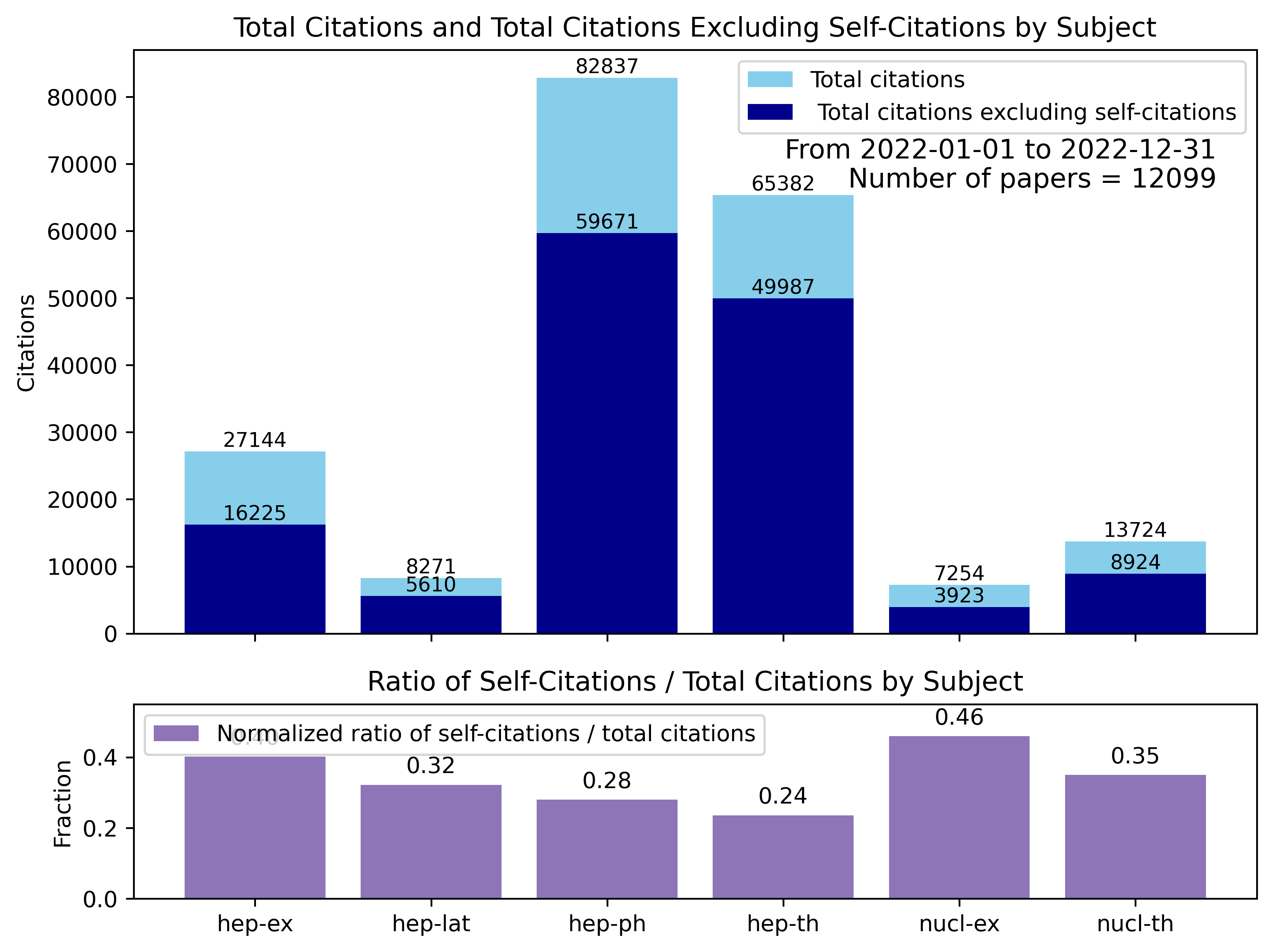}
\label{fig:bibAPI_citations_by_subject_2022}
}
  \caption{(a,b) same as 
  Fig.~\ref{fig:bibAPI_submitted_by_subject}
  and
  Fig.~\ref{fig:bibAPI_citations_by_subject}
  but for the date range 2022 Jan.~01 -- 2022 Dec.~31.}
\label{fig:subject_2022}
\end{figure}

Running the \texttt{BibAPI} script without any arguments,
\begin{verbatim}
$ python ./bibAPI.py
\end{verbatim}
will invoke the default date range
\confirm{(2023-12-04 to 2024-12-05)}
and default subject category \confirm{(hep-ex and nucl-ex)}.
To modify the date range but not default categories, use
\begin{verbatim}
$ python ./bibAPI.py YYYY-MM-DD YYYY-MM-DD
\end{verbatim}
The list of categories cannot be modified
from the terminal without also stipulating
the date range. However,
one can modify the default value of the
\texttt{category} attribute in the
script's \texttt{APIQueryInputs} class.
This \texttt{object} is located near the top of the
\texttt{bibAPI.py} file.

To reproduce the output in Sec.~\ref{sec:examples}, use the command
\begin{verbatim}
$ python ./bibAPI.py 1991-01-01 2025-03-31 
            hep-ex hep-th hep-ph hep-lat
            nucl-ex nucl-th
\end{verbatim}
For the results in App.~\ref{app:extra}, use instead 
\begin{verbatim}
$ python ./bibAPI.py 2022-01-01 2022-12-31 
            hep-ex hep-th hep-ph hep-lat 
            nucl-ex nucl-th
\end{verbatim}
Collecting and processing data for the entire 34-year 
period takes approximately \confirm{36 hours} on a personal laptop.
The seemingly long runtime is due to brief but intentional pauses during 
\textbf{Phase1} and \textbf{Phase2} (see Sec.~\ref{sec:implementation}).
The pauses are needed to avoid (accidental) denial-of-service attacks on 
databases (a security protocol). 
For about \confirm{330k} records, 
\textbf{Phase 1} will write to file 
a list of \textsc{arXiv} identifiers 
 (\texttt{APIQueryInputs.idPath[0]})
that is about \confirm{8 MB}.
The corresponding \textsc{inSpireHEP} record
(\texttt{APIQueryInputs.dataPath[0]}) 
generated by \textbf{Phase2} 
is about \confirm{15 MBs}.
In total, about \confirm{30 MBs} of records and image files are 
written to file.
For the year \confirm{2023}, 
under \confirm{2 MBs} of records and image files are  written to file.
Copies of our datasets are available at
\begin{center}
\href{https://gitlab.cern.ch/riruiz/public-projects/-/tree/master/BibAPI/ResultsFromPaper}{gitlab.cern.ch/riruiz/public-projects/-/tree/master/BibAPI/ResultsFromPaper}
\end{center}

For further help, execute either of the following
\begin{verbatim}
$ python ./bibAPI.py --help  

$ python
>>> import bibAPI
>>> bibAPI.queryHelp()
\end{verbatim}
Both set of commands will print (hopefully) helpful tips.
The command \texttt{dir(bibAPI)} will list 
all ``methods'' and ``attributes'' available in the script 
(technically a \textit{Python} ``module''). 
Likewise, \texttt{help(bibAPI.xxx)} for method/attribute \texttt{xxx}
will print any available documentation.


\section{Additional Results: 1-year run}
\label{app:extra}

In Fig.~\ref{fig:weekday_2022}, 
Fig.~\ref{fig:bibAPI_citations_by_authors_2022},
and Fig.~\ref{fig:subject_2022}
we show the same results as in Sec.~\ref{sec:examples}
but for the smaller data range 
\confirm{2022 Jan.~01--Dec.~31}.
A total of $N_{\rm tot}=12,099$ records were obtained 
from \textsc{inSpireHEP}, with  
$\delta N = 2$ additional manuscripts excluded.
For one year of papers, 
full runtime took \confirm{about 5 hours} 
on a Wi-Fi connection. 
By the end of \textbf{Phase 1}, 
about \confirm{135 MB} of memory was needed. 
By the end of \textbf{Phase 2}, 
memory needs grew to about \confirm{160 MB}. 
\textbf{Phase 4} 
required about \confirm{405 MB} to generate histograms.
Output files required under \confirm{4 MB of storage}.



\end{document}